\documentclass[letterpaper,twocolumn,10pt]{article}
\usepackage{usenix-2020-09}

\usepackage{tikz}
\usepackage{amsmath}
\usepackage{pifont}

\usepackage{filecontents}
\usepackage{booktabs}
\usepackage{tikz}
\usepackage{amsmath}
\usepackage{xspace}
\usepackage{setspace}
\usepackage{filecontents}
\usepackage{enumitem}
\usepackage{algorithm}
\usepackage{algpseudocodex}
\usepackage{multirow}
\usepackage{circledsteps}
\usepackage{wrapfig}
\usepackage{enumitem}
\usepackage{titlesec}
\usepackage{listings}
\usepackage{color}
\usepackage{makecell}
\usepackage{MnSymbol}
\usepackage{circledsteps}
\usepackage{wrapfig}
\usepackage[most]{tcolorbox}
\usepackage[skip=0pt]{caption}
\usepackage[skip=0pt]{subcaption}

\usepackage{titlesec}
\titlespacing*{\paragraph}{0pt}{0.25em}{0.25em}
\titlespacing*{\section}{0pt}{0.5em}{0.5em}
\titlespacing*{\subsection}{0pt}{0.25em}{0.25em}
\titlespacing*{\subsubsection}{0pt}{0.15em}{0.15em}

\definecolor{dkgreen}{rgb}{0,0.6,0}
\definecolor{gray}{rgb}{0.5,0.5,0.5}
\definecolor{mauve}{rgb}{0.58,0,0.82}
\usepackage{relsize}
\usepackage{enumitem}
\usepackage[normalem]{ulem} 
\setenumerate[1]{itemsep=0pt,partopsep=0pt,parsep=\parskip,topsep=0pt,leftmargin=2.2em}
\setitemize[1]{itemsep=0pt,partopsep=0pt,parsep=\parskip,topsep=0pt,leftmargin=2.2em}
\setdescription{itemsep=0pt,partopsep=0pt,parsep=\parskip,topsep=0pt,leftmargin=2.2em}

\newcommand{\projectname}{TENT\xspace}

\begin{document}

\date{}

\title{\Large \bf \projectname: A Declarative Slice Spraying Engine for Performant and Resilient Data Movement in Disaggregated LLM Serving}

\author{
\rm Feng Ren$^{1,2}$, Ruoyu Qin$^{2,3}$, Teng Ma$^{4}$, Shangming Cai$^{4}$, Zheng Liu$^{5,4}$, Chao Lei$^{6}$, Dejiang Zhu$^{6}$, \\ \rm Ke Yang$^{7}$, Zheming Li$^{3}$, Jialei Cui$^{3}$, Weixiao Huang$^{3}$, Yikai Zhao$^{3}$, Yineng Zhang$^{8}$, \\ \rm Hao Wu$^{1}$, Xiang Gao$^{6}$, Yuhao Fu$^{6}$, Jinlei Jiang$^{2}$, Yongwei Wu$^{2}$, Mingxing Zhang$^{2}$ \\
\rm \normalsize $^1$9\# AISoft; $^2$Tsinghua University; $^3$Moonshot AI; $^4$Alibaba Cloud Computing; \\
\rm \normalsize $^5$Zhejiang University; $^6$Ant Group; $^7$Approaching.AI; $^8$Independent Researcher
}

\maketitle

\begin{abstract}


Modern GPU clusters rely on complex, heterogeneous interconnects. As large language model (LLM) serving shifts toward agentic reasoning, KVCache becomes a first-class mobile asset, driving frequent migrations and massive ``elephant flows'' that dominate the execution critical path. Operating Mooncake Transfer Engine (TE) on thousands of GPUs exposed a fundamental flaw in existing frameworks: \textit{imperative, early-binding path selection}. This rigidity results in \textit{state-blind striping} that ignores congestion and grey failures, leading to bandwidth stranding. This also produces \textit{operational fragility} where routine faults require manual intervention.

We present \textbf{TENT}, a \textit{declarative} orchestration engine that decouples transfer intent from physical execution. By abstracting interconnects into a unified resource pool, TENT shifts path resolution from initialization to \textit{slice-time late binding}. Applications simply declare transfer intents, while TENT dynamically ``sprays'' fine-grained slices across rails based on real-time telemetry and predictive cost modeling. This orchestration eliminates head-of-line (HoL) blocking and enables transparent, sub-50~ms self-healing by rerouting slices around failures or degradations without application-level intervention.

TENT serves as the production data plane for LLM inference and reinforcement learning (RL) pipelines at multiple industrial clusters. Our evaluation shows that TENT outperforms state-of-the-art baselines, including Mooncake TE, NIXL, and UCCL. In LLM inference with SGLang HiCache, TENT achieves up to $1.36\times$ higher throughput and $26\%$ lower P90 time-to-first-token (TTFT) than Mooncake TE. In RL pipelines, TENT accelerates parameter updates in Moonshot Checkpoint Engine by $20\sim 26\%$.

\end{abstract}

{
  \renewcommand{\thefootnote}{} 
  \footnotetext{Corresponding author: Mingxing Zhang 
  (\texttt{zhang\_mingxing@mail. tsinghua.edu.cn})}            
  \addtocounter{footnote}{-1}   
}
\section{Introduction}
\label{sec:intro}
LLM serving~\cite{zhong2024distserve,vllm2025,sglang2025,kwon2023efficient,dynamo2025,liu2024deepseek} is transitioning from stateless, single-turn inference to agentic, multi-round reasoning. 
This paradigm shift redefines KVCache from a localized computational byproduct into a first-class mobile asset that must be frequently migrated across heterogeneous nodes to balance compute loads or maintain context~\cite{lmcache2025,qin2025mooncake,wu2026dualpath,liu2025deepseek}.
Similarly, the emergence of online reinforcement learning~\cite{guo2025deepseek,team2025kimi,yao2023deepspeed} necessitates microsecond-scale synchronization of gradients across massive clusters. 
These flows now reside on the execution critical path; any latency jitter in the interconnect is directly compounded across reasoning turns, dictating end-to-end Service Level Objectives (SLOs), particularly TTFT.

\begin{figure}[t]
    \centering
    \includegraphics[width=\linewidth]{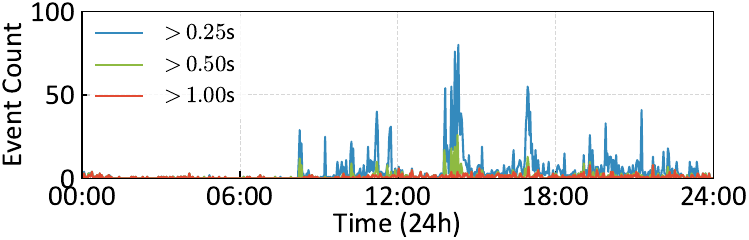}
    \caption{Distribution of the number of end-to-end SLO-violating requests over a 24-hour period in production (08:00--12:00 and 14:00--22:00 are heavily used).}
    \label{fig:ProductionLatency}
\end{figure}

Unlike homogeneous HPC environments, modern AI fabrics feature a hierarchical and fragmented interconnect topology, encompassing PCIe/UPI, multi-rail RDMA, and proprietary protocols (e.g., NVLink~\cite{nvlink2025}, AMD Infinity Fabric~\cite{amd2025}, and Ascend UB~\cite{hixl}).
These links not only exhibit multi-order-of-magnitude performance asymmetry, but also operate under continuous physical churn and diurnal traffic bursts. As quantified in Figure \ref{fig:ProductionLatency}, the number of SLO violations exhibits extreme stochasticity during peak hours. The primary systems challenge has thus shifted from mere point-to-point speed to the \textbf{autonomous orchestration} of these disparate links to maintain resource liquidity in a highly volatile environment.

Production systems rely heavily on advanced P2P transfer engines such as Mooncake Transfer Engine (TE)~\cite{qin2025mooncake,qin2026tos}, NVIDIA NIXL~\cite{nixl2025}, UCCL~\cite{uccl_transport} and DLSlime~\cite{DLSlime2025}. While these systems are masterfully engineered for stable topologies, our operational experience in large-scale heterogeneous clusters reveals a fundamental mismatch: they rely on an \textit{imperative, early-binding model}. By requiring applications to explicitly bind to specific transport backends or physical paths at initialization, this static mapping paradigm struggles in modern dynamic clusters for three systemic reasons:

\noindent\textbf{Imperative early-binding creates communication silos.}
Existing frameworks tether transfer metadata (such as RDMA \texttt{rkey}s~\cite{rdma} or NVLink IPC handles~\cite{cuda}) to specific backends at initialization. This endpoint-centric architecture treats interconnects as isolated domains rather than a fungible resource pool. In production, this rigid coupling precludes the dynamic synthesis of multi-hop paths (e.g., NVIDIA GB200 NVL72 remote DRAM access~\cite{cuda}), mandating brittle manual relay logic and effectively partitioning the fleet.

\noindent\textbf{State-blind scheduling leads to bandwidth stranding.}
Traditional multi-rail aggregation utilizes static hashing or striping, making it state-blind and priority-oblivious. In our deployments, mixing latency-critical MoE (Mixture of Experts)   flows (i.e., mice flows) with throughput-heavy KVCache migrations (i.e., elephant flows) under static striping inevitably triggers HoL blocking on specific degraded rails. This strands healthy auxiliary bandwidth and severely inflates tail latency (P99), particularly during the prolonged diurnal traffic bursts (i.e., 08:00--12:00 and 14:00--22:00, shown in Figure~\ref{fig:ProductionLatency}).

\noindent\textbf{Device-centric design demands manual intervention.} 
Our production telemetry logs an average of 382 failure events per month in large-scale clusters, ranging from GPU dropouts to frequent link flaps. Existing engines treat these as fatal exceptions, throwing errors back to the application layer and forcing manual intervention or expensive control-plane rebuilds. This reactive failure model is fundamentally incompatible with the continuous availability demanded by agentic workloads.

These operational lessons lead to our core thesis: the transfer engine must evolve from a passive, imperative library into an autonomous orchestration plane.
We present \textbf{\projectname}, a declarative transfer framework built for disaggregated LLM serving.
\projectname shifts the paradigm from early physical binding to declarative intent orchestration: applications simply declare \textit{what} data to move and its associated SLOs, while \projectname dynamically resolves \textit{how} to execute it via slice-time late binding across a unified interconnect pool.

Concretely, \projectname makes three key shifts:

\noindent\textbf{From Early Binding to Dynamic Orchestration.} 
    Instead of committing to a specific transport backend at initialization, \projectname provides a platform-agnostic semantic representation for data movement, and leverages multiple transports to interface with diverse interconnects. Central to this shift is a programmable orchestration engine, allowing \projectname to navigate a dynamic reachability map on a per-request basis. This architectural flexibility also enables the autonomous synthesis of transfers via staging host/device memory, effectively bridging vendor-specific ecosystem walls and reducing cross-platform deployment complexity.

\noindent\textbf{From State-Blind Striping to Telemetry-Driven Slice Spraying.} 
    To overcome the HoL blocking inherent in static striping without incurring severe CPU reassembly penalties, \projectname decomposes elephant flows into fine-grained slices (e.g., 64 KB) embedded with absolute destination offsets. It schedules them among all available interconnects based on an adaptive scoring function driven by estimated completion time and real-time telemetry. This single-sided, offset-based addressing natively tolerates out-of-order slice delivery. \projectname also incorporates a priority-aware bandwidth governor that partitions physical link capacity according to application-defined SLOs. This mechanism ensures that even when multiple processes contend for the same physical NIC, high-priority intents receive a guaranteed share of bandwidth and minimal latency.
    
\noindent\textbf{From Manual Intervention to Resilient Self-Healing.} 
    \projectname monitors the link state for each path, and moves error handling from the application into the data plane. It treats transport faults not as exceptions, but as routine routing events: if a path is failed or degraded, the runtime automatically re-schedules it on an alternative path in sub-milliseconds via soft-exclusion and idempotent retries. This \textit{in-band recovery} significantly mitigates transient hardware failures, ensuring reliable delivery without requiring application-level retries or checkpoints.

\projectname has been deployed as the common data‑movement substrate for both inference and reinforcement-learning pipelines in large industrial environments, serving billions of tokens per day on clusters with up to thousands of GPUs.
In a leading financial technology company, \projectname provides explicit SLO guarantees, consistently achieving sub-second TTFT and sub-30 ms TPOT (time-per-output-token) in production.
At a major AI service provider, \projectname handles nearly all production inference traffic, running on a thousand-GPU cluster and processing over 50 million tokens per minute at peak. 

Beyond real-world deployment, we also evaluate \projectname using multiple controlled testbeds to facilitate the effectiveness:

\noindent\textbf{Performance.} Across two production-style workloads, \projectname delivers consistent improvements.
For LLM serving with SGLang~\cite{sglang2025}, it achieves up to $1.36\times$ higher throughput and reduces P90 TTFT by up to $26\%$, accelerating the retrieval of global KVCache blocks.
For RL pipelines with Moonshot Checkpoint Engine~\cite{checkpoint2025}, \projectname accelerates model weight updates by $20\sim26\%$ over Mooncake TE. Host-to-host microbenchmarks on an eight-rail 200~Gbps RDMA fabric confirm that \projectname's slice spraying increases throughput by $33\%$ and reduces P99 latency to $27.6\%$ of the baseline established by Mooncake TE, NIXL, and UCCL-P2P.

\noindent\textbf{Portability.} \projectname runs unmodified across six hardware ecosystems and seven transport protocols. By utilizing lightweight backends ($<800$ LOC), it decouples ecosystem-specific logic while maintaining native performance. This abstraction minimizes engineering overhead for heterogeneous deployment without sacrificing throughput.

\noindent\textbf{Reliability.} \projectname masks both fail-stop failures and soft degradations, restoring throughput within $50$~ms via Telemetry-Driven Slice Spraying and Proactive Dual-Layer Resilience. Validated by a year-long, thousand-GPU production deployment, the engine transparently handles persistent interconnect churn, converting fabric instabilities into minor, transient performance fluctuations.

As a pivotal component of the Mooncake ecosystem, \projectname has been open-sourced at \url{https://github.com/kvcache-ai/Mooncake}. It serves as the high-performance transfer substrate for various subsystems, including Mooncake Store, Mooncake EP, and Mooncake PG, which have progressively integrated \projectname to achieve superior data movement efficiency.



\section{Operational Case Studies with Mooncake TE}
\label{sec:case-study}
\label{sec:interconnects}
\begin{figure}[t]
    \centering
    \includegraphics[width=0.8\linewidth]{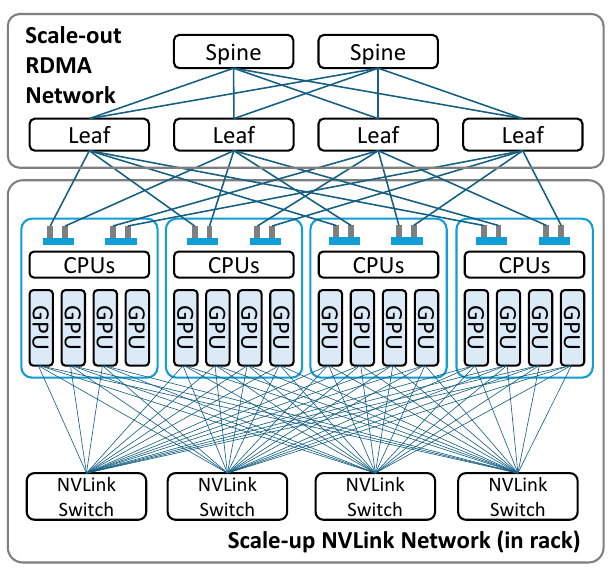}
    \caption{Hierarchical interconnect architecture in a modern GPU supernode. }
    \label{fig:intra-network}
\end{figure}

Existing communication engines (e.g., UCX, NIXL, NCCL) adhere to an \textit{imperative path-selection model} where applications commit to a fixed backend stack at compilation or initialization. They assume that: (1) connectivity is a static, transitive property of an endpoint pair; (2) path selection can be resolved via hard-coded heuristics at initialization; and (3)~the application manages the connection lifecycle.

As illustrated in Figure~\ref{fig:intra-network}, our operations of a massive, disaggregated LLM serving fleet proves these assumptions no longer hold. The hierarchical coexistence of scale-up NVLink networks and scale-out RDMA networks introduces structural boundaries that static models cannot traverse. As hardware diversity increased, the rigidity of this model manifested as significant operational bottlenecks. This section details three representative challenges that motivated the design of \projectname.

\subsection{Communication Silos}
As clusters scale into thousands of GPUs, this early-binding approach creates fragmented communication silos by failing to adapt to the fluid nature of modern interconnects.

\paragraph{Deficiency in Metadata Representation.}
While existing engines offer unified APIs, they function as thin wrappers that mask deep metadata silos. Applications are still forced to manage backend-specific descriptors, such as RDMA memory keys (\texttt{ibv\_reg\_mr})~\cite{rdma} for the scale-out fabric or NVLink IPC handles (\texttt{cudaIpcMemHandle} or \texttt{cuMemExportToShareableHandle})~\cite{cuda} for the scale-up fabric, which remain mutually unintelligible across platform boundaries. Even when engines like UCX~\cite{ucx2025} support both RDMA and NVLink, such integration is often ecosystem-locked (e.g., NVIDIA-only) and lacks the vendor-neutrality required for truly heterogeneous fleets.

This semantic incompatibility creates communication silos, forcing operators to manage multiple independent engine instances to leverage disparate resources. To resolve this, \projectname introduces a unified metadata abstraction and runtime orchestration, leveraging late-binding to decouple application logic from hardware-specific drivers and enable seamless mixed-deployment.

\paragraph{Partial Fabric Coverage.}
The second challenge is that high-speed fabrics are functionally constrained to specific data movement patterns. As illustrated in Figure~\ref{fig:intra-network}, the scale-up NVLink network is optimized for low-latency GPU-to-GPU memory copies. However, traditional engines like UCX~\cite{ucx2025} rely on static heuristics that fail when a transfer intent deviates from these rigid specifications.
We observed a representative case in a production environment featuring only a scale-up fabric. The engine failed to utilize the NVLink fabric for some access patterns (e.g., transferring KVCache from remote DRAM, which is \textit{not} natively supported)~\cite{cuda}. Because the communication intent did not match the hard-coded direct-link criteria, the system executed a fallback to the TCP-based RPC. This resulted in severe performance degradation while the specialized scale-up interconnect remained underutilized.

To address this challenge, autonomous orchestration is essential to dynamically provision appropriate staging buffers, enabling data transfers to leverage high-speed interconnects that would otherwise be inaccessible.


\subsection{Bandwidth Stranding}
\label{sec:study-2}
The operational performance of disaggregated serving depends on the efficient coordination of conflicting data flows over a shared interconnect backplane. The bandwidth stranding problem arises because existing engines rely on static, state-blind path selection, which fails to utilize available capacity due to non-uniform fabric load.

\paragraph{Mixed-traffic Interference.}
The interconnect substrate simultaneously accommodates latency-critical MoE Expert Parallelism (\textit{mice flows}) and throughput-heavy KVCache migrations (\textit{elephant flows}). 
Current practices mitigate mixed-traffic interference by rigidly partitioning workloads—typically confining MoE traffic to intra-node NVLink while reserving RDMA for KVCache migrations. Such static isolation not only limits MoE scaling beyond the single-node boundary but also leads to systemic resource under-utilization. In scenarios where resource sharing is unavoidable, such as cross-node MoE EP flows, this state-blindness frequently stalls latency-sensitive, critical-path signals behind elephant flows, effectively stranding aggregate bandwidth even when complementary physical paths remain idle.

\paragraph{Overhead from Straggler.}
\begin{figure}[t]
    \centering
    \includegraphics[width=\linewidth]{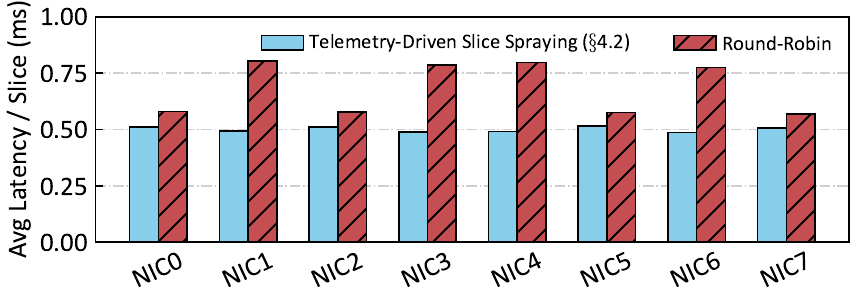}
    \caption{Per-rail average latency. TENT eliminates the HoL blocking spikes observed in Round-Robin (RR).}
    \label{fig:scheduling-problem}
    \vspace{-0.5em}
\end{figure}

The cache-to-compute ratio of 22 GB/PFLOP in models such as DeepSeek-V3 identifies KVCache data movements characterizing the critical execution path of agentic reasoning tasks~\cite{wu2026dualpath,liu2025deepseek}. Our production telemetry also reveals that large-scale KVCache transfers frequently suffer from state-blind striping in existing engines (e.g., UCX-backend NIXL and Mooncake TE). By distributing chunks via static hashing or NUMA priorities, these engines fail to account for grey failures and noisy neighbors common in production. Assigning slices to such degraded or cross-NUMA rails creates stragglers, inducing HoL blocking that stalls the entire transfer.

We reproduced this behavior experimentally on an eight-rail 200 Gbps RDMA fabric. We issued 1~MB read requests with four submission threads. We compared the round-robin baseline against the telemetry-driven scheduler (\S\ref{sec:adaptive-scheduling}). Under round-robin, rails attached to remote NUMA domains exhibited significantly higher per-slice service times; the resulting queue buildup on those specific rails inflated P99 latency for the entire request, as shown in Figure~\ref{fig:scheduling-problem}. In contrast, when the scheduler utilized live telemetry (queued bytes and completion latency) to steer slices away from backlogged rails, throughput increased, tail latency dropped, and progress remained balanced across the fabric.

Thus, on the same heterogeneous fabric that exposed static binding issues, state-blind striping turned physical non-uniformity into performance cliffs. \projectname addresses this by treating all rails as a unified resource pool, decomposing transfers into fine-grained slices, and scheduling each slice dynamically based on a telemetry-aware cost model.


\subsection{Fabric Churn and Manual Recovery}
\label{sec:study-3}

\begin{table}[t]
\centering
\smaller
\caption{Datacenter failure breakdown collected over a one-month period.}
\begin{tabular}{l r r}
\hline
\textbf{Failure Event} & \textbf{Count} & \textbf{\%} \\
\hline
\textbf{GPU}: ECC Errors & 154 & 40.2 \\
\textbf{GPU}: Device Dropout & 92 & 24.2 \\
\textbf{GPU}: XID Errors & 12 & 3.2 \\
\textbf{GPU}: Device Enumeration Failures & 9 & 2.4 \\
\textbf{GPU}: Over-Temperature Events & 10 & 2.5 \\
\hline
\textbf{Node}: Crashes & 30 & 7.9 \\
\textbf{Node}: Motherboard / PCIe / BMC Failures & 15 & 3.9 \\
\hline
\textbf{Network}: Cable Fault & 15 & 3.8 \\
\textbf{Network}: Frequent Link Down Events & 6 & 1.6 \\
\textbf{Network}: NIC Hardware Failures & 4 & 1.0 \\
\hline
Others & 35 & 9.3 \\
\hline
\textbf{Total} & \textbf{382} & \textbf{100.0} \\
\hline
\end{tabular}
\label{tab:failure-classification}
\vspace{-0.5em}
\end{table}

The third class of issues arose from ordinary hardware churn in large datacenters. Even in well-maintained clusters, components fail and recover continuously. In one representative deployment at a leading fintech company, we observed on average 382 failure events per month with a mean repair time of 160.21 minutes, shown in Table~\ref{tab:failure-classification}. 

These disturbances often appear as rail-level instabilities, including both hard faults (e.g., NIC flapping) and soft failures like Priority Flow Control (PFC) storm. In engines using static binding, such localized issues induce HoL blocking, where a single slow or failing rail ``pollutes'' the entire multi-rail fabric and constrains the aggregate throughput. This necessitates dynamic NIC combinations, enabling the data plane to utilize real-time telemetry to re-spray slices across alternative paths, effectively isolating localized glitches.

Furthermore, GPU-centric errors account for over 70\% of the observed failure events (Table~\ref{tab:failure-classification}), which usually requires replacing the node. While monolithic models like NCCL~\cite{nccl2025} rely on global state synchronization—where the failure of a single device often triggers a cluster-wide restart, a P2P-based transfer engine should treat this as a dynamic membership event rather than a fatal exception. By decoupling transfer tasks from the global cluster state, the engine is designed to minimize the fault scope and ensure that localized glitches do not compromise overall system availability, thereby avoiding the global re-initialization.

In summary, \projectname shifts the recovery paradigm from error absorption to prevents localized churn from escalating into the cascading failures or global cold starts. Its two-layer resilience architecture embeds health monitoring and backend substitution into the data plane to mask transient instabilities. Simultaneously, the control plane manages long-term availability through asynchronous membership updates. 


\section{\projectname Architecture}
\label{sec:overview}


\projectname transitions the transfer engine from a passive, imperative library into an \textbf{autonomous orchestration plane}. The core of this architecture is the \emph{Declarative Intent} interface. Unlike prior systems (e.g., Mooncake TE or UCX) that rely on connection-time binding of hardware backends, \projectname decouples the application's logical requirements from the engine's physical realization.

\begin{figure}[t]
    \centering
    \includegraphics[width=\linewidth]{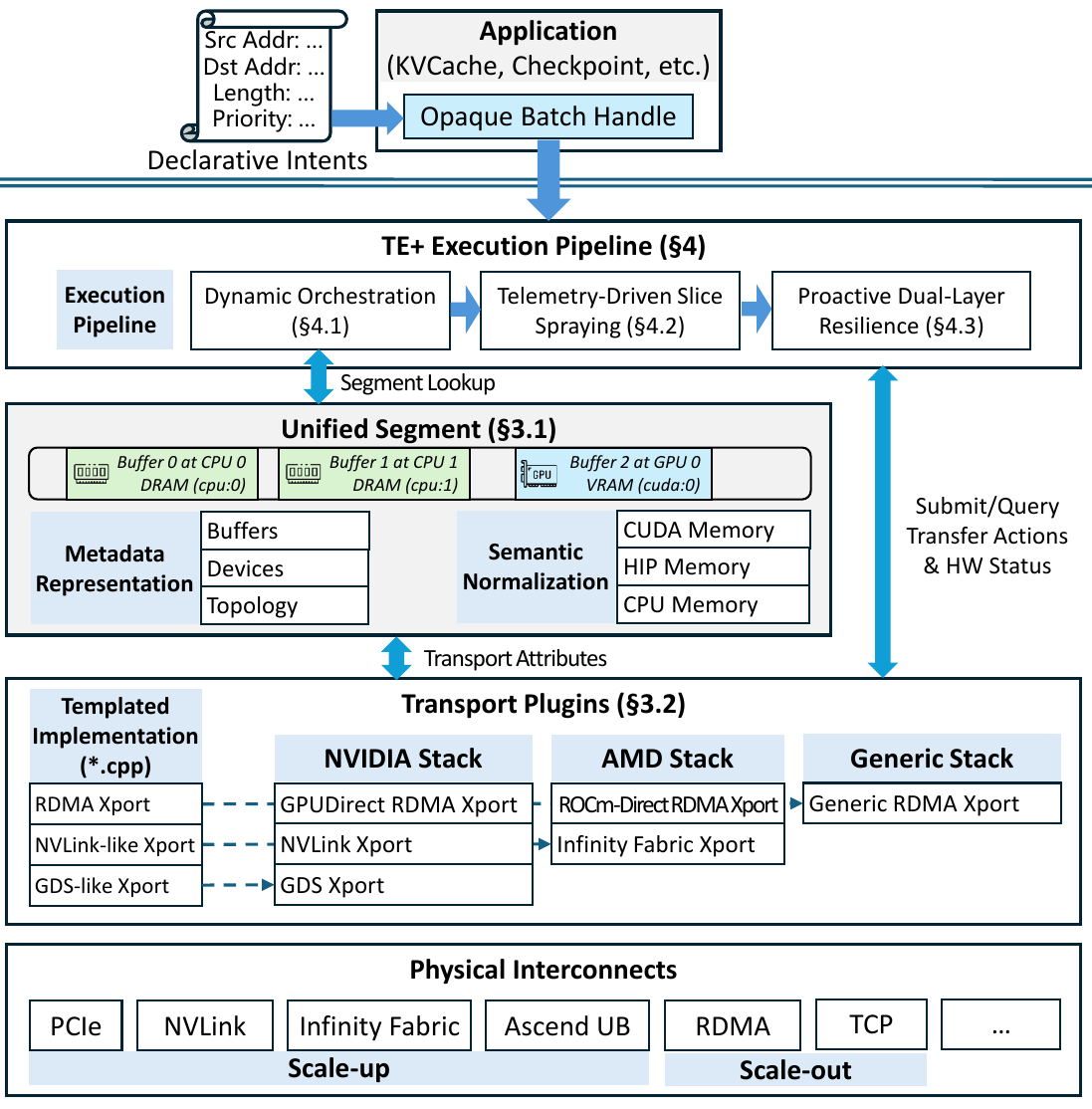}
    \caption{\projectname's architecture and execution pipeline.}
    \label{fig:modules}
    \vspace{-1em}
\end{figure}

Through this interface, applications submit high-level intents (e.g., KVCache migration or checkpointing) and receive an \textit{opaque batch handle} (Figure~\ref{fig:modules}). This handle encapsulates the movement of data between \textit{logical segments} without exposing transport-specific descriptors or explicit endpoints. 
By offloading routing, resilience, and load balancing to the data plane, \projectname eliminates the {manual orchestration burden} across heterogeneous clusters.
To realize this declarative paradigm, \projectname organizes its internal architecture as shown in Figure~\ref{fig:modules}. It reconciles intents via a three-phase {execution pipeline} (\S\ref{sec:pipeline}).

\subsection{Unified Segment Representation}
\label{sec:intent}

To bridge the {communication silos} created by fragmented storage and interconnects, \projectname\ introduces the unified segment abstraction. A segment normalizes heterogeneous memory domains—spanning GPU/NPU HBM, host DRAM, and NVMe-oF—into a single, location-agnostic namespace. By reconciling mutually unintelligible device descriptors (e.g., RDMA \texttt{rkeys} or CUDA IPC handles) into a normalized metadata context, segments decouple the application’s logical intent from its physical realization.

\begin{figure}[t]
    \centering
    \includegraphics[width=\linewidth]{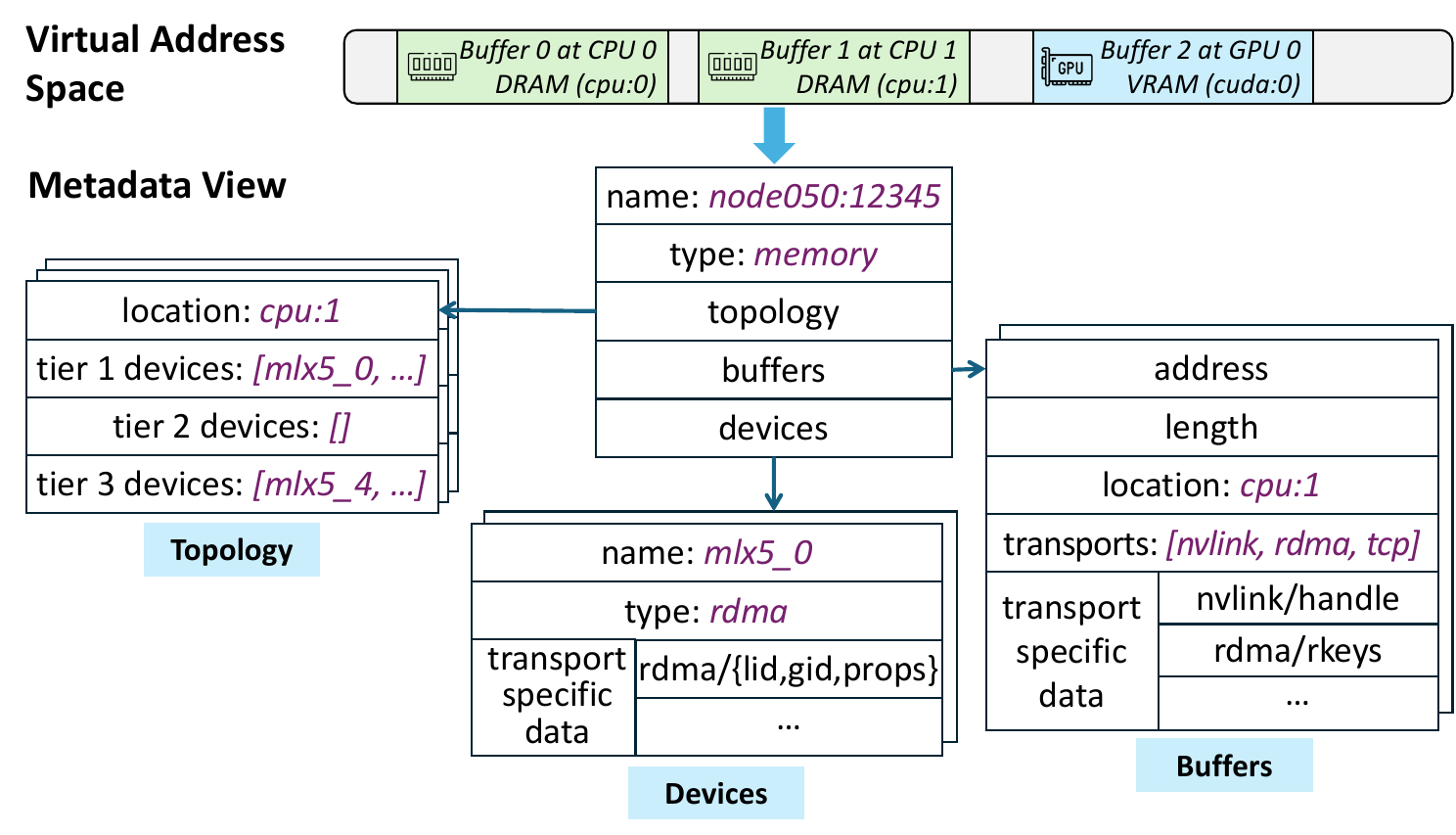}
    \caption{Metadata representation of a memory segment.}
    \label{fig:segment-format}
\end{figure}

As illustrated in Figure~\ref{fig:segment-format}, this abstraction is materialized as a hierarchical metadata view. Each segment is identified by a unique logical name and encapsulates three core sub-structures: \texttt{Topology}, \texttt{Devices}, and \texttt{Buffers}. Crucially, the transport-specific descriptors remain opaque to the core engine, stored within the \textit{transport-specific data} fields. This strict isolation ensures that \projectname can autonomously resolve cross-fabric reachability and integrate emerging hardware without affecting the upper-layer orchestration logic.

\paragraph{Building Segment Metadata.}
At initialization, \projectname\ performs automated topology discovery to populate these hierarchical structures. It enumerates all available resources and classifies their interconnects into protocol-independent affinity tiers: \textbf{tier-1} for native high-speed paths (e.g., NVLink), \textbf{tier-2} for cross-PCIe-root connections, and \textbf{tier-3} for NUMA-crossing fallbacks.
This information is encoded into the segment's topology field, defining the physical location and reachability (shown in Figure~\ref{fig:segment-format}). Furthermore, the buffers sub-structure maintains a curated \texttt{transports} list, identifying valid communication protocols for that specific memory type. By exposing these affinities, \projectname provides the necessary state for the execution pipeline to query the transport-level capability matrix and perform \textit{dynamic path synthesis} (\S\ref{sec:orchestration}).

\subsection{Pluggable Transport Plugins}
\label{sec:backends}
To execute data movement across heterogeneous fabrics, \projectname defines a uniform slice-execution interface implemented as runtime-loadable plugins. As illustrated in Figure~\ref{fig:modules}, these plugins encapsulate vendor-specific memory APIs and asynchronous hardware primitives. By unifying these details under a standardized {slice-execution interface}, \projectname can route transfer tasks across multi-vendor hardware without requiring conditional logic in the orchestrator.

Thanks to many GPU vendors already provide CUDA-compatible APIs, \projectname achieves code reuse across for different platforms. Supporting a new platform requires only defining a semantic normalization, i.e., pointers to some functions related to device memory operations. Furthermore, since the core orchestrator remains strictly platform-agnostic, \projectname achieves single-binary portability. The engine dynamically loads the appropriate transport plugins (\texttt{.so}) based on the detected environment at runtime, eliminating the need for vendor-specific container images and simplifying cross-architecture deployment.

The transport plugins also facilitate slice late-binding through a standardized \textit{capability profile} (e.g. allowing device-to-device inter-node transfer), so that \projectname can decide whether a request can utilize a high-speed direct path (e.g., NVLink) or trigger path synthesis (\S\ref{sec:orchestration}) to compose multiple fabrics via staging buffers. This makes \projectname completely masking hardware heterogeneity from the application.

\section{\projectname Execution Pipeline}
\label{sec:pipeline}

Based on the declarative abstractions, \projectname implements a three-phase execution pipeline (Figure~\ref{fig:modules}) to transform logical intents into physical data movement: dynamic orchestration (\S\ref{sec:orchestration}) chooses paths at request time, telemetry-driven slice spraying (\S\ref{sec:adaptive-scheduling}) distributes large transfers across healthy rails, and dual-layer resilience (\S\ref{sec:resilience}) mitigates link and backend failures. We also outline datapath optimizations that make these mechanisms practical at scale (\S\ref{sec:put-together}).

\subsection{Phase 1: Dynamic Orchestration}
\label{sec:orchestration}

\projectname implements transport selection at the submission time of each transfer request. Unlike UCX~\cite{ucx2025}, which requires applications to explicitly manage stateful endpoints, \projectname abstracts these details behind a declarative interface, deferring the mapping of logical intents to internal transport plugins until \texttt{submitTransfer}. Upon submission, the engine returns an opaque batch handle (Figure~\ref{fig:modules}), which serves as the unique identifier for the entire lifecycle of the transfer, regardless of the underlying physical complexity.

\paragraph{Transport Resolution.}
Instead of burdening applications with the manual management of stateful endpoint handles, \projectname delegates transport resolution to the orchestrator at request time. Upon receiving a transfer intent, the orchestrator performs a stateless resolution by intersecting source and destination metadata (\S\ref{sec:intent}) with the capability profile (\S\ref{sec:backends}) of available transport plugins. This process evaluates (1) physical reachability via heterogeneous fabrics, (2) topological affinity tiers (e.g., PCIe vs. NVLink domains), and (3) request-specific QoS constraints. By dynamically resolving the execution plan at the moment of dispatch, \projectname ensures that data movement is always directed to the optimal available path, effectively shielding application logic from underlying connection churn and fabric reconfigurations.

Instead of relying on static heuristics only, \projectname employs a rule-based engine to reconcile automated topology sensing with user-defined predicates. This allows administrators to define high-level policies—such as isolating high-priority KVCache flows to dedicated RDMA rails or mandating NVLink usage for latency-critical metadata. By fusing the ``ground truth'' of discovered hardware affinity with declarative application intents, the orchestrator minimizes the operational coordination required to achieve performance isolation in multi-tenant environments.

\begin{figure}[t]
    \centering
    \includegraphics[width=\linewidth]{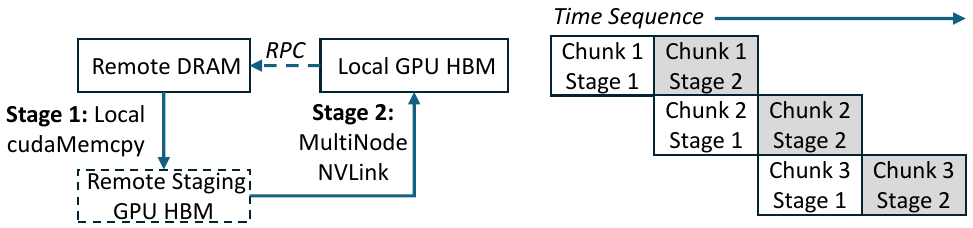}
    \caption{Example of Autonomous Path Synthesis: reading data from remote DRAM via scale-up network only.}
    \label{fig:Path-Synthesis}
    \vspace{-1em}
\end{figure}

\paragraph{Autonomous Path Synthesis.}
In scale-up supernodes (e.g., NVIDIA GB200 NVL72)~\cite{nvlink2025}, direct P2P connectivity is often constrained by fabric addressability limits, creating communication silos. \projectname addresses these reachability gaps through a {compositional transfer paradigm}. When no direct path exists between segments, the orchestrator autonomously synthesizes a multi-stage execution pipeline by stitching together discrete transport primitives. For instance, a transfer from remote host DRAM to a local GPU is decomposed into a  sequence (as shown in Figure~\ref{fig:Path-Synthesis}).

The implementation of this synthesis focuses on maximizing hardware utilization through {fine-grained staging and overlapping}. The orchestrator allocates intermediate buffers within the local NUMA domain to minimize cross-die latency and partitions the payload into pipelined chunks. This allows the system to overlap the execution of consecutive stages—for example, the RDMA transfer of the $n$-th chunk into local DRAM occurs concurrently with the PCIe DMA transfer of the $(n-1)$-th chunk into GPU HBM. By masking the complexity of multi-hop routing behind a single completion event, \projectname provides a unified elastic buffer abstraction that normalizes fragmented storage into a continuous high-speed namespace.

\subsection{Phase 2: Telemetry-Driven Slice Spraying}
\label{sec:adaptive-scheduling}
While the previous phase determines the logical route, the second phase focuses on physical execution efficiency. \projectname decomposes large transfers into fine-grained slices and schedules each slice to the rail expected to complete it fastest based on live telemetry.

\paragraph{Slice Decomposition.}
For large requests, payloads are partitioned into slices with a minimum size of 64~KB. This granularity is a calculated trade-off: it is small enough to mitigate HoL blocking on congested rails, yet large enough to amortize the CPU overhead of doorbell rings and polling. Each slice may be issued as an independent work request that writes directly to its final destination offset, enabling out-of-order completion without CPU-side reordering.

\begin{algorithm}[t]
\smaller
\caption{Telemetry-Aware Slice Scheduling}
\label{alg:predict-feedback}
\begin{algorithmic}[1]
\Require Request length $L$, target node $b$, priority $p$, threshold $\theta$
\Ensure Set of scheduled tasks $\mathcal{T} = \{ \langle \pi, \text{offset}, \text{length} \rangle \}$

\Statex \textbf{// Stage 1: QoS-Based Path Filtering}
\State $\mathcal{D}_{admitted} \gets \{ d \in \text{AvailableDevices} \mid \text{CheckQoS}(d, p) = \text{True} \}$
\If{$\mathcal{D}_{admitted} = \emptyset$} \Return \textsc{Defer\_Execution} \EndIf

\Statex \textbf{// Stage 2: Predictive Cost Modeling and Path Binding}
\State $\mathcal{P} \gets \emptyset$ \Comment{Set of candidate path triples $\langle \pi_i, s_i \rangle$}
\For{each device $d^{loc}_{i} \in \mathcal{D}_{admitted}$}
    \State $d^{rem}_{i} \gets \text{FindRemoteDevice}(d^{loc}_{i}, b)$
    \State $\pi_i \gets \langle d^{loc}_{i}, d^{rem}_{i} \rangle$ \Comment{Paired NICs represent a path}
    \State $\hat{t}_i \gets \beta_{0,i} + \beta_{1,i} \cdot \frac{A_i + L}{B_i}$ \Comment{Predicted cost for a base slice}
    \State $s_i \gets P_i \cdot \hat{t}_i$ \Comment{Apply topology penalty $P_i$}
    \State $\mathcal{P} \gets \mathcal{P} \cup \{ \langle \pi_i, s_i \rangle \}$
\EndFor
\State $s_{min} \gets \min \{ s_i \mid \langle \pi_i, s_i \rangle \in \mathcal{P} \}$ \Comment{Minimal prediction complete time}

\Statex \textbf{// Stage 3: Mode-Specific Traffic Dispatching}
\If{$L < \theta$} \Comment{Mode A: Mice flows (Latency-sensitive)}
    \State $\mathcal{C} \gets \{ \pi_i \mid \langle \pi_i, s_i \rangle \in \mathcal{P} \land s_i \le (1+\gamma) \cdot s_{min} \}$
    \State $\pi^* \gets \text{RoundRobin}(\mathcal{C})$ 
    \State \Return $\{ \langle \pi^*, L \rangle \}$ \Comment{Dispatch request to path $\pi^*$}
\Else \Comment{Mode B: Elephant flows (Throughput-optimized)}
    \State \Return $\text{WeightedSpray}(L, \mathcal{P})$
\EndIf
\end{algorithmic}
\end{algorithm}

\noindent\textbf{Predictive Cost Modeling.}
To inform scheduling, \projectname employs a predictive cost model to estimate the completion time $\hat{t}_d$ for a slice of length $L$ on device $d$:
\begin{equation}
\hat{t}_d = \beta_{0,d} + \beta_{1,d} \cdot \frac{A_d + L}{B_d}
\label{eq:pred-time}
\end{equation}
where $A_d$ is the effective queue length (bytes currently in-flight) and $B_d$ is the estimated \textit{link bandwidth}. In this model, $\beta_{0,d}$ and $\beta_{1,d}$ act as conservative system-level coefficients that absorb the software-stack overheads, such as memory registration checks and asynchronous submission latencies.

A critical challenge in operational environments is distinguishing transient congestion from permanent link degradation. Currently, \projectname implements a \textit{peak-picking} strategy to infer the link bandwidth by isolating application-level load. Within each measurement window, the system identifies the cleanest path condition:
\begin{equation}
B^{\text{probe}}_d = \min \left\{ B^{\text{hw}}_d, \max_i \frac{\alpha \cdot L_i}{\tau_i - \delta} \right\}
\label{eq:bw-probe-qual}
\end{equation}
where $B^{\text{hw}}_d$ is the theoretical bandwidth reported by hardware, $L_i$ and $\tau_i$ denote the data length and end-to-end latency of the $i$-th slice, respectively. Here, $\delta$ and $\alpha$ are environment-specific calibration constants configured to align the raw probe observations with the operational envelope of the target fabric. To ensure scheduling stability, the capacity $B_d$ is updated via a statistical smoothing filter (e.g., EWMA) applied to the observed $B^{\text{probe}}_d$ peaks. 

\noindent\textbf{Slice Dispatching.}
As illustrated in Algorithm~\ref{alg:predict-feedback}, scheduling involves slice dispatching in two steps.

In \textit{Path Binding}, \projectname prioritizes {routing stability} by enforcing a topology-aligned mapping that matches affinity tiers between source and destination (e.g., intra-PCIe domain). This fixed-route strategy localizes traffic within optimal physical domains, minimizing switch hops and fabric pressure. While favoring stable paths, the binding is {dynamically resilient}: the orchestrator autonomously fails over to alternative NICs if primary endpoints degrade. A topology penalty $P_d$ (e.g., 0.1 for same-NUMA vs. 5.0 for cross-NUMA) is applied to the cost $s_d$, ensuring the system only deviates from optimal routing when necessary to sustain aggregate throughput.

The second stage performs \textit{Slice Dispatching}. For small transfers, \projectname selects paths within a cost tolerance to introduce stochastic diversity. For elephant flows, the system employs weighted proportional spraying, distributing slices according to efficiency weights $W_i$. This approach prevents herd behavior by statistically dispersing traffic across all feasible rails, ensuring the system converges toward optimal aggregate utilization.

\begin{figure}[t]
    \centering
    \includegraphics[width=0.85
    \linewidth]{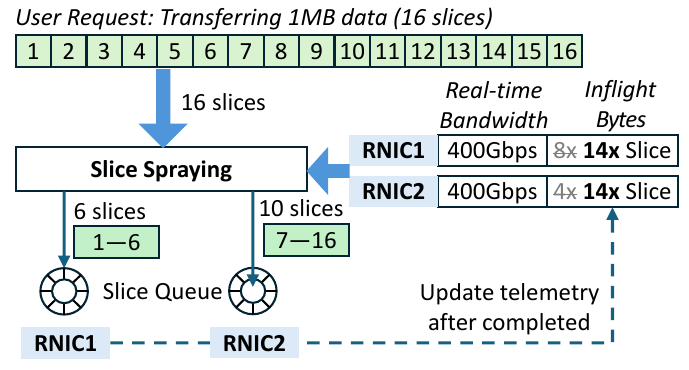}
    \caption{Spraying 1MB data (16 slices) among two devices.}
    \label{fig:Slice-Spraying}
    \vspace{-1em}
\end{figure}

\noindent\textbf{Quality of Service.} 
To ensure SLO-awareness in shared clusters, \projectname implements a two-tier QoS framework. Locally, it utilizes priority-slot rotation, where devices cycle through \texttt{HIGH}, \texttt{MID}, and \texttt{LOW} priority windows. A request only consumes a device's bandwidth if its priority $p$ satisfies the current slot's threshold. Globally, per-device queue depths are synchronized via shared memory, preventing noisy-neighbor interference where background data migrations might otherwise starve latency-sensitive inference tokens. This hardware-software co-design ensures that critical SLO targets are met regardless of the background transfer volume.

\subsection{Phase 3: Proactive Dual-Layer Resilience}
\label{sec:resilience}
The third phase focuses on maintaining high availability under fabric churn (\S\ref{sec:study-3}). By operating at both the rail and transport layers, \projectname ensures that transient hardware issues are resolved in-band, maintaining a stable completion event for the application.

\paragraph{Health Monitoring.}
At the link level, \projectname monitors both implicit and explicit signals. The telemetry loop from Phase 2 naturally detects struggling rails as their predicted completion times grow relative to peers; in addition, backends surface explicit errors such as RDMA completion failures, timeouts, or GPU-side events. When a rail crosses a degradation threshold, the scheduler applies soft exclusion by effectively setting its cost to infinity and removing it from the candidate set, without invoking heavyweight reconfiguration. A background prober periodically sends lightweight heartbeat slices to excluded rails; once a rail proves responsive and stable, its cost is reset and it is gradually re-admitted into the scheduling pool. 

More specifically, the rail monitor tracks each local--remote NIC pair across three health states: (1) \texttt{HEALTHY}: normal operation; (2) \texttt{DEGRADED}: elevated latency or packet loss, where the rail receives a 0.5$\times$ weight penalty but remains eligible for traffic; and (3) \texttt{PAUSED}: error threshold exceeded, leading to temporary exclusion with an exponential backoff cooldown. To prevent unnecessary path switching during transient load spikes, \projectname distinguishes between congestion-related errors (e.g., \texttt{RNR\_RETRY\_EXC}) and hard hardware failures. Congestion errors require twice the default error threshold before triggering failover, ensuring the system remains robust against micro-bursts.
Both \texttt{DEGRADED} and \texttt{PAUSED} states support automatic recovery to maximize resource utilization. A \texttt{DEGRADED} rail is promoted to \texttt{HEALTHY} after a period without degradation events. For \texttt{PAUSED} rails, \projectname employs a doubling cooldown period before the rail re-enters the candidate set.

\paragraph{Link-Level Resilience.}
When a slice fails due to a hard error or timeout, \projectname executes an idempotent retry mechanism. Each slice writes to an absolute destination offset, making retries safe even if data was partially written but the completion notification was lost. To identify the most reliable alternative, \projectname employs a two-pass path rotation strategy based on the current retry count. In the first pass, the scheduler attempts to find a \texttt{HEALTHY} rail, rotating through available source devices to avoid localized congestion. If no healthy paths are available, it falls back to \texttt{DEGRADED} rails, prioritizing affinity-matched (same-NUMA) pairs before attempting any reachable remote NIC.
Although these retries bypass the predictive cost model, their resource consumption is still included in the global queue statistics used by Phase 2, so recovery traffic does not starve unrelated flows.

\paragraph{Transport-Level Resilience.}
The {declarative decoupling} of \projectname enables an ultimate safety net: backend substitution. Unlike early-bound systems where backend failures necessitate application-level re-initialization, \projectname's orchestrator can transparently reroute slices across heterogeneous fabrics (e.g., failing over from RDMA switches). This architectural agility ensures that even catastrophic backend failures manifest only as transient performance jitters rather than service interruptions, directly fulfilling the {goal-fulfillment} philosophy of the engine.

\subsection{Low-Overhead Datapath}
\label{sec:put-together}
To ensure orchestration logic does not become a bottleneck, \projectname implements a zero-copy, asynchronous datapath designed for high-concurrency slice processing. 

\projectname decouples logical intent from hardware execution through a lock-free, multi-producer single-consumer (MPSC) ring buffer. When an application invokes \texttt{submitTransfer}, the orchestrator resolves the transport plan and pushes slice descriptors into the ring buffer, allowing the application thread to return immediately. This design prevents application-level stalls and shields the control plane from hardware-specific doorbell latencies.

Pinned worker threads asynchronously dequeue slices, employing opportunistic batching to minimize submission overhead. By aggregating multiple slices into a single linked-list work request (e.g., a multi-WR \texttt{ibv\_post\_send}), \projectname significantly reduces CPU cycles per byte. Each worker owns a dedicated transport context (e.g., RDMA Queue Pair), eliminating doorbell contention. Furthermore, \projectname leverages one-sided hardware primitives for direct-to-offset writes, enabling fully zero-copy, CPU-bypass reception that remains robust against out-of-order slice arrival.

To scale to millions of outstanding slices, \projectname replaces fine-grained polling with a hierarchical atomic counter mechanism. Worker threads poll completion queues in batches and decrement an atomic counter within the corresponding batch's control block for each completed slice. Applications only observe coarse-grained batch completion, minimizing cache-coherence traffic and synchronization overhead. This lightweight completion path ensures the datapath sustains near-line-rate throughput while supporting \projectname's advanced scheduling and resilience mechanisms.


\section{Evaluation}
We evaluate \projectname across diverse, production-representative GPU clusters to verify its capability in securing latency SLOs for agentic workloads, masking hardware churn, and providing uniform performance across fragmented fabrics.

\paragraph{Testbed and Baselines.}
Our evaluation is conducted on a heterogeneous fleet mirroring large-scale industrial deployments:
\begin{itemize}
    \item \textbf{H800:} Two nodes, each equipped with $8\times$ NVIDIA H800 GPUs and $8\times$ 200~Gbps RoCE NICs.
    \item \textbf{H20:} Four nodes, each equipped with $8\times$ NVIDIA H20 GPUs and $4\times$ 400~Gbps RoCE NICs.
    \item \textbf{Others:} To evaluate portability across fragmented hardware, we include NVIDIA GB200 NVL72 clusters and Ascend clusters with UB/HIXL fabrics.
\end{itemize}


Unless otherwise noted, we compare \projectname against three production-level P2P engines: Mooncake TE (\textbf{TE})~\cite{qin2025mooncake}, our production predecessor; \textbf{NIXL}~\cite{nixl2025}, NVIDIA's UCX-based cross-NIC transfer framework; and \textbf{UCCL-P2P}~\cite{uccl_transport}, the P2P backend in the UCCL communication framework. All engines run on the same hardware and RDMA configuration. Mooncake TE and UCCL-P2P are configured with their recommended multi-rail setups (uniform striping or per-region NIC pinning), while NIXL uses its default UCX policy, which selects two best NICs based on static transport properties.

\subsection{End-to-End Study}

\subsubsection{KVCache-intensive LLM Serving}
\label{sec:kvcache}

To evaluate \projectname's ability to handle intensive KVCache retrieval in agentic reasoning, we integrate it into \textbf{SGLang HiCache}~\cite{sglang2025}. We evaluate \projectname using a 10-turn agentic conversation benchmark with the Qwen3-235B-A22B-Instruct-2507  model~\cite{yang2025qwen3} on the H800 cluster, with model weights sharded via tensor parallelism ($TP=8$) per 8-GPU node. To simulate production-level concurrency, 60 clients issue interactive requests (2048 input tokens each), creating bursty, inter-node KVCache migrations—typical elephant flows on the execution critical path. We compare \projectname against Mooncake TE and a non-cached baseline, using a 600~GB KVCache budget for all caching variants.

Table~\ref{fig:hicache_result} summarizes the results. Compared to the non-cached baseline, \projectname delivers a $3.79\times$ throughput boost and an $83.4\%$ reduction in P90 TTFT. Crucially, the benefits amplify as conversation length grows: by round 10, \projectname maintains a sub-second average TTFT ($0.66$~s), whereas the baseline exceeds $4$~s. More importantly, \projectname outperforms Mooncake TE by $1.36\times$ in throughput and $26.4\%$ in P90 TTFT. These gains stem entirely from our transport-layer optimizations, as all caching policies remain identical.

The performance gap highlights the limitation of imperative, early-binding models. While Mooncake TE treats interconnects as isolated silos and relies on state-blind RDMA striping, \projectname elevates NVLink as a first-class transport and employs slice-time late binding. We prioritize intra-node NVLink paths for local slices while aggregating multi-rail RDMA for inter-node transfers. The resulting fabric-aware data movement translates directly into lower TTFT and higher throughput for KVCache retrieval workloads.

\paragraph{Operational Results.}
At a major AI service provider, \projectname handles nearly all production inference traffic, running on a thousand-GPU cluster and processing over 50 million tokens per minute at peak. In agentic API services, where multi-turn reasoning relies heavily on context reuse, \projectname maintains a typical $90\%$ cache hit rate in high-concurrency production. This efficiency reduces the effective billing cost for users to just $25\%$ of the standard market price, significantly sharpening the commercial competitiveness of large-scale LLM offerings.

\begin{table}[t]
\caption{Multi-turn conversation benchmark of SGLang Hi-Cache using \projectname and Mooncake TE.}
\smaller
\centering
\label{fig:hicache_result}
\begin{tabular}{l|p{1.1cm}|p{1.7cm}|p{0.8cm}}
\hline
\textbf{Metric}      & \textbf{Baseline} & \textbf{Mooncake TE} &\textbf{\projectname} \\ \hline
Input Throughput (tokens/s) & 20,757 & 58,006 & 78,759 \\
\hline
Average TTFT (s) &  2.12 & 0.72 & 0.53       \\ 
P90 TTFT (s) &  4.02 & 0.90 & 0.67 \\ 
\hline
R1 Avg TTFT (s) &  0.38 & 0.45 & 0.43       \\ 
R5 Avg TTFT (s) &  1.89 & 0.68 & 0.52       \\ 
R10 Avg TTFT (s) &  4.09 & 0.97 & 0.66      \\ 
\hline
\end{tabular}
\end{table}

\subsubsection{Reinforcement Learning Parameter Updates}
\label{sec:checkpoint}

We evaluate \projectname in the context of high-frequency model weight updates in online RL pipelines, where the refresh latency should be short enough to avoid stalling sample rollouts. 
We use \textbf{Moonshot Checkpoint Engine v0.2.0}~\cite{checkpoint2025, team2025kimi}, a middleware for in-place model weight updates that supports both cluster-wide broadcast via NCCL and fine-grained P2P weight transfers through a pluggable backend (Mooncake TE or \projectname).
We run the engine on the H800 cluster ($TP=8$, FP16 weights), where model updates manifest as massive, synchronized elephant flows so that all ranks participate in data transfer during each update.

We compare Mooncake TE and \projectname as backends while keeping the checkpoint format, sharding, and update schedule fixed. For each model, we measure the end-to-end checkpoint apply time: the interval between initiating the update and all ranks installing the new weights. Table~\ref{fig:p2p_performance} shows that \projectname reduces the update time by $19.7\%$ and $26.1\%$ for Qwen3-235B-A22B-Instruct-2507 and GLM-4.5-Air~\cite{zeng2025glm} models respectively. 
\projectname dynamically orchestrates intra-node NVLink and inter-node RDMA. By spraying slices based on real-time link pressure, it effectively mitigates the ``straggler'' effect caused by non-uniform PCIe-NUMA topologies.

\paragraph{Operational Results.}
To evaluate scalability, we integrate the Moonshot Checkpoint Engine with \projectname and deploy it on a $256\times$ H20 (TP=16, 32 nodes) semi-production cluster.
Across trillion-parameter models including DeepSeek-V3.1~\cite{liu2024deepseek} and Kimi-K2-Instruct~\cite{team2025kimi}, \projectname consistently synchronizes model weights across hundreds of GPUs within tens of seconds. This predictable, high-bandwidth performance enables frequent, zero-downtime parameter refreshes, allowing model updates to proceed without impacting service availability.

\begin{table}[t]
\caption{Parameter update time (s) with Moonshot Checkpoint Engine on an $8\times$H800 (TP8) FP16 testbed. }
\smaller
\centering
\label{fig:p2p_performance}
\begin{tabular}{l|l|l}
\hline
\textbf{Model}      & \textbf{Mooncake TE} & \textbf{\projectname} \\ \hline
Qwen3-235B-A22B-Instruct-2507 &  12.87 & 10.34       \\ 
GLM-4.5-Air &  7.17 & 5.30  \\ \hline
\end{tabular}
\end{table}

\begin{figure*}[t]
    \centering
    \includegraphics[width=\linewidth]{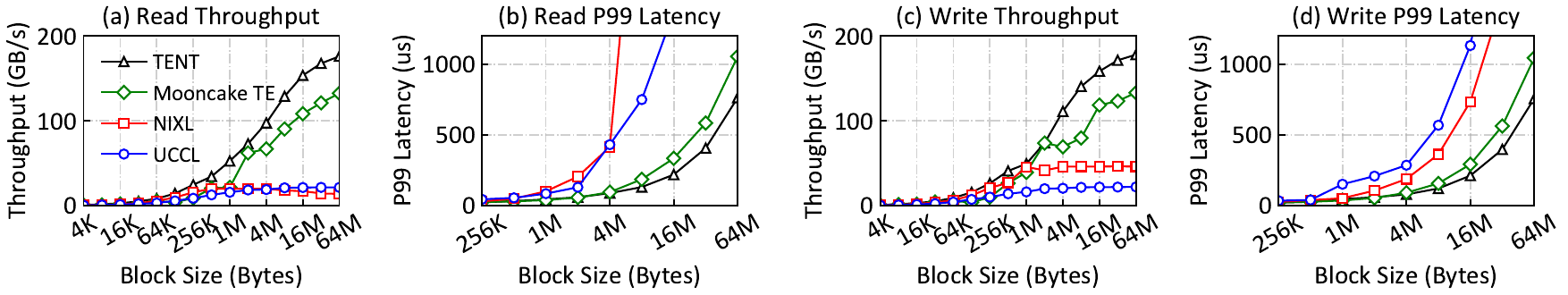}
    \caption{Host-to-host read/write throughput and P99 latency between two H800 nodes, with memory allocated per socket and transfers issued by per-socket threads.}
    \label{fig:host-to-host}
    \vspace{-1.5em}
\end{figure*}

\subsection{Performance}
We then present microbenchmarks to show that \projectname exploits multi-rail fabrics more effectively than existing engines across a wide range of block sizes and concurrency levels, and they explain the end-to-end improvements observed in LLM serving and checkpoint update workloads.

\begin{figure}[t]
    \centering
    \includegraphics[width=\linewidth]{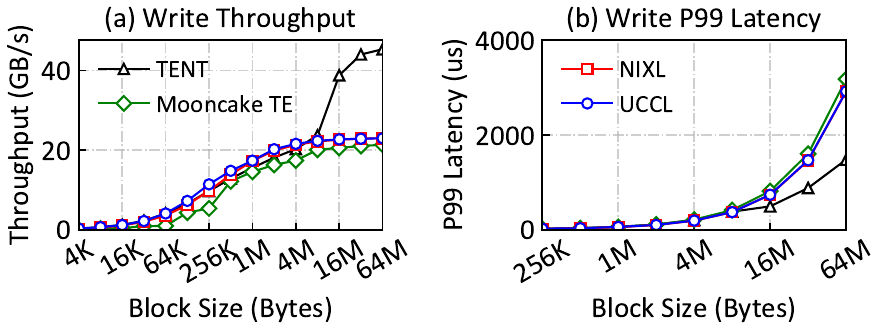}
    \caption{Write throughput and P99 latency for point-to-point transfers between GPUs on two H800 nodes.}
    \label{fig:gpu-to-gpu}
\end{figure}

\subsubsection{Microbenchmarks}
To isolate data-plane efficiency from framework overheads, we utilize \textbf{TEBench} (mocking NIXLBench~\cite{nixl2025}) to evaluate \projectname's scheduling and multi-rail orchestration. For NIXL and UCCL-P2P~\cite{uccl_transport}, we use their native benchmarking scripts to avoid measurement bias.

\paragraph{Host-to-Host.}
We evaluate host-to-host transfers between two H800 nodes using eight 200 Gbps RoCE NICs. For each CPU socket, we allocate pinned memory and one submission thread (batch size 1, 4 KB to 64 MB blocks). While both Mooncake TE and \projectname use all available rails, \projectname improves write throughput by $33.7\%$, and reduces P99 latency by $27.6\%$. \projectname consistently outperforms NIXL and UCCL-P2P for blocks $\ge$ 1 MB.

The performance gap stems from rail-level slice scheduling. UCCL-P2P binds memory to a single NIC without aggregation, capping throughput at per-NIC limits. NIXL defaults to two NICs and stripes transfers via static bandwidth rankings. Both lack real-time queue visibility; consequently, transiently overloaded rails bottleneck throughput and dominate P99 latency. \projectname avoids these issues by tracking rail queue depth and completion behavior to detect congestion missed by static policies. By assigning slices based on observed service time, \projectname redirects flows from slow rails before queues accumulate, maintaining high aggregate bandwidth and reducing tail latency.

\paragraph{Device-to-Device.}
We also evaluate GPU-originated flows where PCIe and NUMA topologies introduce significant rail asymmetry. Each H800 GPU in our testbed has one tier-1 NIC with the closest PCIe affinity, and three tier-2 NICs on the same NUMA node. Benchmarking inter-node writes with typical KVCache block sizes shows that \projectname improves throughput by $2.1\times$ and reduces P99 latency to $46.7\%$ relative to Mooncake TE (Figure~\ref{fig:gpu-to-gpu}).

Unlike UCCL-P2P and Mooncake TE, which fix mappings to a single NIC, \projectname evaluates predicted service times across all rails using topology-dependent penalties. While the tier-1 NIC remains dominant for small blocks, \projectname recruits tier-2 NICs for large blocks once parallel bandwidth gains outweigh access penalties. Per-NIC byte counters confirm this, showing roughly half the data offloaded to tier-2 rails. By selectively adding rails only when they reduce total service time, \projectname effectively supports the high-bandwidth requirements of KVCache and model-update transfers.

\begin{figure}[t]
    \centering
    \includegraphics[width=\linewidth]{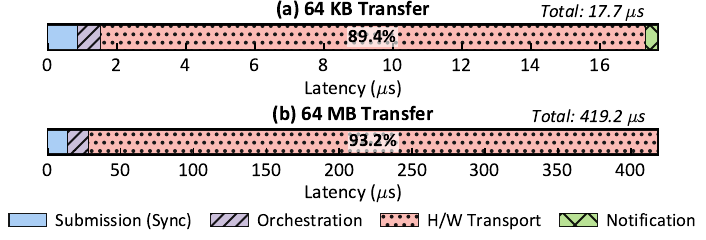}
    \caption{Latency breakdown of submitting 64KB and 64MB requests on H20 nodes.}
\label{fig:latency_breakdown}
\end{figure}

\paragraph{Breakdown.}
To evaluate the overhead of our declarative orchestration, we decomposed the end-to-end latency for 64 KB and 64 MB transfers on H20 nodes. In our execution model, synchronous blocking is restricted exclusively to the submission phase. As illustrated in Figure~\ref{fig:latency_breakdown}, a 64 KB slice transfer incurs a synchronous software overhead of only $1.5\mu\text{s}$, allowing the user thread to be released immediately while subsequent orchestration proceeds asynchronously. For a 64 MB payload, the end-to-end latency is $419.2\mu\text{s}$, and hardware transport accounts for $93.2\%$ of total latency. Consequently, the system is fundamentally wire-bound, and \projectname control plane does not bottleneck high-concurrency data movements.

\subsubsection{Concurrency Scaling}

We study how \projectname behaves under increasing request-level parallelism, as in multi-tenant serving nodes, by varying the number of submission threads and batch sizes.

\begin{figure}[t]
    \centering
    \includegraphics[width=\linewidth]{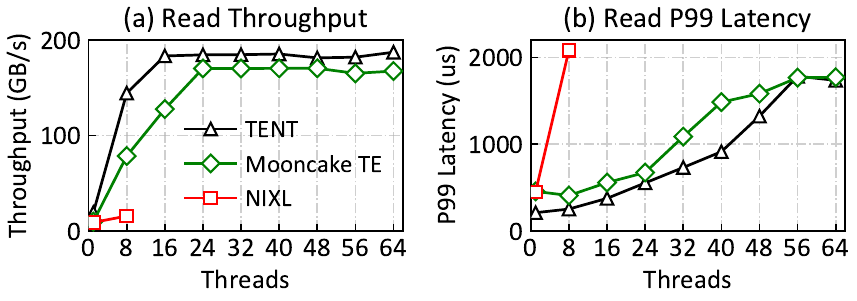}
    \caption{GPU-to-GPU read performance measured with varied submission threads on H800, with block size 4~MB.}
    \label{fig:threads}
\end{figure}


\paragraph{Submission Thread.}
We scale submission threads from 1 to 64 and measure GPU-to-GPU read bandwidth using 4 MB blocks, with each thread bound to a local GPU (Figure \ref{fig:threads}). When all eight GPUs issue transfers, \projectname sustains 144 GB/s—over $2\times$ Mooncake TE and about $77\%$ of hardware peak. Notably, \projectname saturates bandwidth with only 16 threads and maintains a wide margin over NIXL even as concurrency grows. This efficiency stems from two properties: \projectname actively steers slices away from congested rails, preventing the hot-rail stalls that constrain Mooncake TE, and it achieves high throughput with few backend workers, minimizing submission and polling contention. As a result, \projectname reaches peak bandwidth without requiring large thread counts, making it significantly more robust and friendly to elephant flows.

\paragraph{Batch Size.}
\begin{wrapfigure}{r}{0.5\linewidth} 
  \centering
  \vspace{-4mm} 
  \includegraphics[width=\linewidth]{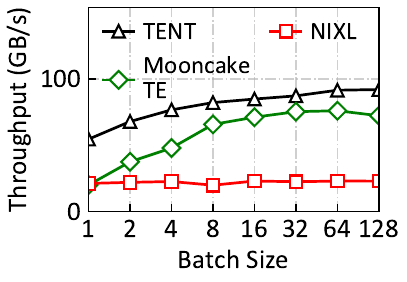}
  \vspace{-2mm} 
  \caption{Host-to-host write throughput with 4 NICs and varied batch size.}
  \label{fig:batch}
\end{wrapfigure}
We vary the batch size from 1 to 128 using a single submission thread and 4 MB blocks, with source and destination buffers pinned to host memory on NUMA node 0 (Figure~\ref{fig:batch}). Given the four local NICs, the aggregate theoretical bandwidth is 800 Gbps. In this configuration, NIXL fails to aggregate bandwidth and utilizes only a single NIC because 4 MB blocks fall below its static multi-rail activation threshold. In contrast, \projectname approximates the hardware line rate as the batch size increases, delivering $1.16\times$ to $2.72\times$ the bandwidth of Mooncake TE and reducing P90 latency by an average of $27.06\%$. While Mooncake TE relies on state-blind randomized selection—frequently bottlenecked by the slowest rail—\projectname explicitly incorporates queued bytes and completion telemetry to mitigate straggler effects.

\subsubsection{Quality of Service}
We evaluate \projectname's priority-aware bandwidth governor on the H20 testbed using two concurrent processes, each employing 8 submission threads to saturate the $4\times$ 400~Gbps RoCE fabric. The first process issues latency-sensitive ``mice flows'' (64~KB metadata sync, always in high priority), while the second generates throughput-heavy ``elephant flows'' (64~MB KVCache migrations). We compare four configurations: \textit{No QoS}, \textit{QoS (High+Medium)}, \textit{QoS (High+Low)}, and the \textit{Solo} execution baseline. High-priority intents utilize priority-slot rotation and low-priority traffic is actively suppressed to maintain low physical queue depth.

\begin{figure}[t]
    \centering
    \includegraphics[width=\linewidth]{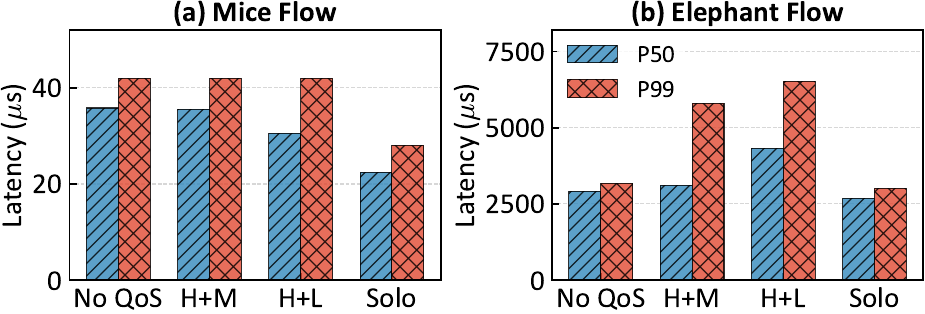}
    \caption{Performance isolation of co-located workloads under different QoS policies: P50/P99 latencies of $64\text{ KB}$ mice flows (High Priority) and $64\text{ MB}$ elephant flows (Medium/Low Priority).}
    \label{fig:qos-impact}
\end{figure}

Results in Figure~\ref{fig:qos-impact} demonstrate a 15.1\% reduction in mice-flow P50 latency, primarily achieved by mitigating inter-process contention through intentional time-slot throttling. This is enabled by the engine's priority-slot rotation, which prioritizes high-priority slices while ensuring consistent forward progress to prevent starvation. Notably, P99 latency remains stable because it accepts all slices regardless of priority when a device rotates to the ``accept-low'' state, ensuring that high-priority intents are never indefinitely stalled behind background traffic. These results confirm that \projectname's QoS-aware slice spraying effectively prioritizes latency-critical mice flows (such as MoE Expert Parallelism), directly contributing to the reduction of overall TTFT.

\subsection{Portability}

\begin{table}[t]
    \centering
    \caption{Peak and theoretical read bandwidth across different transfer modes. $\dagger$ indicates peak bandwidth reported by vendors or measured by native benchmarks.}
    \smaller
    \begin{tabular}{l|r|r}
        \hline
        \textbf{Transport} & 
        \makecell{\textbf{Measured}\\\textbf{(GB/s)}} & 
        \makecell{\textbf{Theoretical}\\\textbf{(GB/s)}$^{\dagger}$} \\
        \hline
        RDMA: GPU$\rightarrow$GPU (H800) & 44.9 & 25.0/rail \\
        RDMA: GPU$\rightarrow$GPU (H20) & 45.7 & 50.0 \\
        - RDMA: GPU$\rightarrow$GPU (Target Staging) & 46.2 & 50.0 \\
        - RDMA: GPU$\rightarrow$GPU (Initiator Staging) & 33.6 & 50.0 \\
        - RDMA: GPU$\rightarrow$GPU (Dual Staging) & 33.2 & 50.0 \\
        \hline
        \texttt{io\_uring}: GPU$\rightarrow$File & 6.0 & 6.0 \\
        NVLink: GPU$\rightarrow$GPU & 172.0 & 204.5 \\
        MNNVL: GPU$\rightarrow$GPU & 781.8 & 956.2 \\
        Ascend: GPU$\rightarrow$GPU & 135.0 & 196.0 \\
        \hline
    \end{tabular}
    \label{tab:peak-bandwidth}
\end{table}

\projectname provides a unified declarative interface that masks interconnect complexity while maintaining near-native performance. We evaluate its portability across the following dimensions.

\paragraph{Engineering Overhead.}
Currently, \projectname supports six heterogeneous ecosystems with minimal code overhead, including NVIDIA CUDA (26 LOCs), AMD HIP (28), Ascend (130), Moore Threads MUSA (92), MetaX MACA (75), and CPU-only (125). It further integrates seven transport protocols to ensure fleet-wide connectivity, spanning high-performance RDMA (6,975 LOCs) and NVLink (531 for intra-node and 691 for inter-node) to specialized backends like Ascend Direct (904), NVIDIA GPUDirect Storage (481), and TCP-based RPC (322). By decoupling ecosystem-specific logic from data-plane execution, our transport protocols remain generic across all supported hardware. This modularity significantly reduces code duplication and facilitates streamlined updates, allowing protocol optimizations to be deployed across the entire heterogeneous fleet.

\paragraph{Runtime Overhead.}
A central design goal is to expose a unified declarative interface across heterogeneous fabrics while incurring negligible overhead. We evaluate peak bandwidth across diverse transports to quantify how close \projectname comes to theoretical hardware limits and to measure the efficiency of its autonomous path synthesis in resolving reachability gaps.

Across all tested fabrics, \projectname tracks native performance closely (Table~\ref{tab:peak-bandwidth}). For intra-node H800 NVLink GPU-to-GPU transfers, it reaches $172.0$~GB/s, and matches native performance without measurable indirection cost. Similar efficiency is observed on high-performance proprietary fabrics: \projectname achieves $781.8$~GB/s on multi-node NVLink of GB200 NVL72, and $135.0$~GB/s on Ascend Supernode. For storage-intensive tasks, the engine attains 6.0~GB/s over \texttt{io\_uring}, saturating the hardware limit. In all cases, applications issue identical \texttt{BatchTransfer} calls, while the engine dynamically resolves the execution backend.

\paragraph{Autonomous Path Synthesis.}
Beyond multi-node NVLink, autonomous path synthesis is critical for interconnecting GPUs lacking native GPUDirect RDMA. We evaluate this by disabling GPUDirect RDMA on both sides, forcing \projectname to transfer via intermediate host memory. As shown in Table~\ref{tab:peak-bandwidth}, Target-side staging mode achieves 46.2~GB/s, matching direct P2P RDMA throughput (45.7~GB/s). This confirms that fine-grained pipelining masks intermediate memory copy overhead by overlapping execution across transport segments. Even with dual-side staging (where initiator-side PCIe is constrained), \projectname sustains 33.2~GB/s. Our declarative orchestration thus normalizes fragmented hardware clusters into a high-speed namespace, providing near-native performance where direct P2P is unavailable.

\subsection{Reliability}
We finally evaluate whether \projectname's dual-layer resilience turns common fabric failures into short-lived performance perturbations rather than application-visible errors. 

\paragraph{Failure Injection in Testbed.}
\begin{figure}[t]
    \centering
    \begin{subfigure}{\linewidth}
        \centering
        \includegraphics[width=\linewidth]{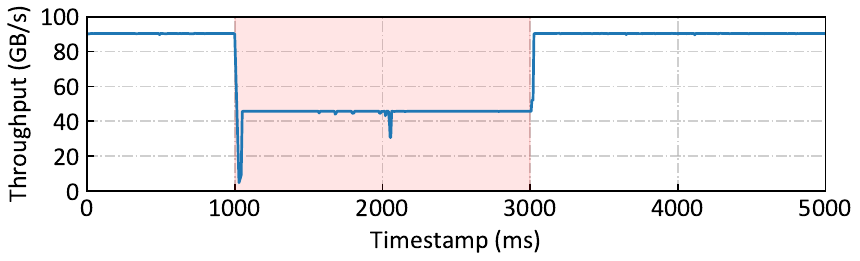}
        \caption{Aggregate bandwidth with a single rail fail-stop and recovery.}
        \label{fig:fail_stop}
    \end{subfigure}
    \hfill
    \begin{subfigure}{\linewidth}
        \centering
        \includegraphics[width=\linewidth]{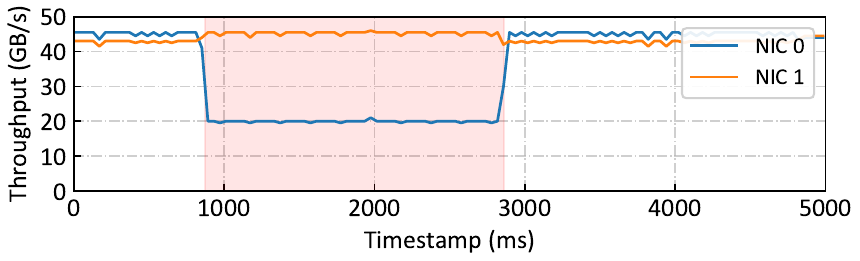}
        \caption{Per-rail bandwidth with NIC 0 degradation and recovery.}
        \label{fig:soft_degradation}
    \end{subfigure}
    \caption{Resilience of \projectname under network anomalies.}
    \label{fig:resilience_comparison}
    \vspace{-1em}
\end{figure}

To evaluate resilience, we inject controlled NIC failures on the H800 testbed during continuous $64$~MB transfers. At $t = 1000$~ms, we force a failure on NIC 0 and recover it at $t = 3000$~ms (Figure~\ref{fig:fail_stop}). Upon failure, \projectname marks the rail unavailable, reroutes subsequent slices, and re-executes aborted ones on healthy rails. This transition causes a throughput dip of less than 50~ms. Periodic health checks every second (longer in production) produce minor fluctuations during path probing. At $t = 3000$~ms, \projectname reintegrates the restored NIC within $26$~ms, returning throughput to nominal levels. All retries and substitutions are handled internally, ensuring zero application-visible failures. These results demonstrate that \projectname converts fabric failures into transient performance blips and recovers routes automatically without application intervention.

Furthermore, as shown in Figure~\ref{fig:soft_degradation}, we evaluate \projectname's response to partial link rate degradation, a common occurrence due to thermal throttling or optical signal interference. When we artificially constrain NIC 0's throughput to half of its nominal capacity, the scheduler dynamically recalibrates the traffic distribution weights. The system enters a new steady state where NIC 0 operates at its degraded capacity (from $\sim 45$~GB/s down to $\sim 20$~GB/s) without causing head-of-line blocking for the entire transfer. Upon lifting the constraint, the transmission rate returns to full line speed within $40$~ms, validating the sensitivity and stability of our EWMA-based dynamic load-balancing algorithm.

\paragraph{Operational Results.}
\projectname has served as the primary data plane for a thousand-GPU cluster for over a year, handling nearly all production inference traffic for a major AI provider. Throughout this deployment, the engine has transparently managed persistent interconnect churn (including per-rail slowdowns, transient link flaps, and device resets), maintaining stable performance without exposing transport exceptions to applications.

\section{Related Works}
\paragraph{LLM Communication Frameworks.}
\textit{Collective communication libraries} like NCCL~\cite{nccl2025,hu2025demystifying}, RCCL~\cite{rocm2025}, HCCL~\cite{uccl_transport}, and others~\cite{dubey2024llama,shah2025msccl,gloo2025} extend group primitives to GPUs. While efficient for training, they assume homogeneous, static topologies and are often vendor-coupled, limiting their use in disaggregated serving.
In contrast, \textit{Point-to-point} (P2P) frameworks such as Mooncake TE~\cite{qin2025mooncake}, NIXL~\cite{nixl2025}, DLSlime~\cite{DLSlime2025}, NVSHMEM~\cite{nvshmem2025}, and UCCL-P2P~\cite{uccl_transport} provide direct memory access suited for dynamic workloads like prefill-decode disaggregation~\cite{zheng2024sglang, sglang2025,vllm2025,kwon2023efficient}, KVCache storage~\cite{lmcache2025,cheng2025lmcache,chen2025impress}, and RLHF~\cite{team2025kimi}.
However, as mentioned in \S\ref{sec:case-study}, existing P2P engines typically rely on static transport binding (e.g., fixed RDMA or NVLink paths) and lack runtime adaptation.
Separately, \textit{Expert Parallelism} (EP) frameworks~\cite{liu2024deepseek,uccl_transport} couple communication with computation to optimize MoE flows, prioritizing kernel-level efficiency over the unified abstraction of diverse heterogeneous interconnects.

\paragraph{Transport Orchestration.}
UCX~\cite{ucx2025} offers unified memory-to-memory abstractions over RDMA, TCP, and SHM, selecting paths using established priorities at connection setup. Systems such as NIXL~\cite{nixl2025} and DLSlime~\cite{DLSlime2025} unify remote memory and storage but generally rely on application-directed backend choices. GeminiFS~\cite{qiu2025geminifs} offers direct file-based access to NVMe
storage managed by the host file system.
In contrast, \projectname adopts a declarative interface that decouples intent from execution, enabling dynamic per-request path selection across heterogeneous fabrics, which is beneficial for disaggregated LLM serving where link conditions evolve over time.
While Beluga~\cite{yang2026beluga} demonstrates the superiority of CXL 2.0 in reducing TTFT by 89.6\% over Mooncake, the declarative abstraction of \projectname remains invariant across evolving backends, allowing for the seamless incorporation of CXL as a pluggable transport plugin.

\paragraph{Transport-Level Performance Optimization.}
While P2P engines (e.g., Mooncake TE~\cite{qin2025mooncake}, NIXL with UCX~\cite{nixl2025,ucx2025}) support multi-rail RDMA, they often rely on static strategies like round-robin. \projectname improves this via telemetry-driven slice scheduling (\S\ref{sec:adaptive-scheduling}).
Prior works~\cite{kalia2016design,wei2021characterizing,dragojevic2014farm,ren2024scaling} have optimized low-level RDMA efficiency (e.g., doorbell batching).
Research also identifies performance bottlenecks in multi-tenant NICs~\cite{kong2023understanding, wang2023srnic} and addresses scale-out QP scalability challenges via techniques like eRPC~\cite{kalia2019datacenter}, Flock~\cite{monga2021birds}, and SMART~\cite{ren2024scaling}. \projectname leverages these low-level insights within a higher-level declarative runtime. 

\paragraph{Fault Tolerance.}
Most existing frameworks rely on reactive fault handling.
UCX~\cite{ucx2025} uses callbacks to report failures, requiring the application to explicitly close and recreate invalid endpoints.
NCCL~\cite{nccl2025} supports asynchronous detection (AER) but halts ongoing operations, requiring global communicator reinitialization.
NIXL~\cite{nixl2025} lacks explicit recovery mechanisms, deferring error handling to underlying transports.
In contrast, \projectname integrates dual-layer resilience, enabling proactive link-level failover and transparent transport-level recovery without user intervention.

\section{Conclusion}
TENT overcomes the operational rigidity of modern GPU clusters by shifting data movement from imperative binding to declarative slice spraying. By unifying heterogeneous interconnects into a dynamic, self-healing resource pool, TENT transparently optimizes multi-rail bandwidth and masks hardware churn. Validated at production scale, it delivers up to $1.36\times$ higher serving throughput and $26.1\%$ faster model updates, demonstrating that a resilient, declarative data plane is essential for next-generation disaggregated AI infrastructure.

\bibliographystyle{plain}
\bibliography{references}

\end{document}